\documentclass[twocolumn,preprintnumbers,amsmath,amssymb]{revtex4}

\usepackage{graphicx}
\usepackage{dcolumn}
\usepackage{bm}

\begin{document}

\title{Constraints on the Physical Parameters of the Dark Energy \\
Using A Model-Independent Approach}

\author{Ruth A. Daly}
 \email{rdaly@psu.edu}
\affiliation{%
Department of Physics, Pennsylvania State University, Berks Campus, Reading, PA 19610\\
}%

\author{S. G. Djorgovski}
 \email{george@astro.caltech.edu} 
\affiliation{
Division of Physics, Mathematics, and Astronomy, 
California Institute of Technology, MS 105-24, Pasadena, CA 91125\\
}%

\date{December 21, 2005}

\begin{abstract}

Understanding the physical nature of the dark energy which
appears to drive the accelerated expansion of the unvierse
is one of the key problems in physics and cosmology today. 
This important problem is best studied using a variety 
of mutually complementary approaches.
Daly and Djorgovski (2003, 2004) 
proposed a model independent approach to determine a number
of important physical parameters of the dark energy as
functions of redshift  directly from the data.  Here, we expand 
this method to include the determinations of its  
potential and kinetic energy as functions of redshift.    
We show that the dark energy potential and kinetic
energy 
may be written as combinations of the first 
and second derivatives of the coordinate distance with respect to redshift.
We expand the data set to include new supernova measurements, and now use
a total of 
248 coordinate distances that span the redshift range from 
zero to 1.79.  First and second derivatives of the coordinate distance
are obtained as functions of redshift, and these are combined to determine 
the potential and kinetic energy  of the dark energy 
as functions of redshift.  An update on the redshift behavior of the
dimensionless expansion rate $E(z)$, the acceleration rate $q(z)$, and the 
dark energy 
pressure $p(z)$, energy density $f(z)$, and equation of state $w(z)$ is also
presented.   We find that the standard $\Omega_{0m}  = 0.3$ and
$\Omega_{\Lambda} = 0.7$ model is in an excellent
agreement with the data.  We also show tentative evidence that 
the Cardassian and Chaplygin gas models in a spatially flat universe 
do not fit the data as well.

\end{abstract}

\maketitle

\section{\label{sec:level1}Introduction}

The acceleration of the universe at the present epoch can
be studied using a variety of techniques such as 
type Ia supernovae \cite{R98, P99, 
T03, K03, B04, R04, A05}, and 
studies of cosmic microwave background radiation
(e.g., \cite{S03}) combined with large scale structure
studies (e.g. \cite{US05,AS05}), 
which indicate
that at the present epoch 
the universe is expanding at an accelerting rate. 
Studies of powerful FRII radio galaxies 
also indicate that the universe is accelerting
\cite{GDW00, DG02, PDMR03}.
To study this in detail, it is important
to determine the redshift behavior of the 
expansion and acceleration rates 
of the universe, and the properties 
and redshift evolution of the driver(s)
of these rates.  

In this vein,  
\cite{dd03,dd04} suggest a model-indepedent method of 
studying the redshift evolution of the expansion and acceleration rates 
of the universe, and show that 
the acceleration
of the universe can be written in terms
of the first and second derivatives of the coordinate distance
with respect to redshift.  
These derivatives are obtained directly from the data on 
coordinate distances using a statistically robust numerical
technique.  
The acceleration thus derived requires
very few assumptions, relying only upon the assumptions that the
universe is homogeneous and isotropic on large scales, and 
has zero space curvature. Indeed, this approach allows a determination
of the acceleration of the universe that is independent of
a theory of gravity and of the properties and redshift
evolution of the drivers of the expansion and acceleration, 
such as dark energy, dark matter, or other components.  
Complementary  model-independent approaches 
based on an integral rather than a differential
technique
are discussed
by \cite{HT99, HT01, SRSS00, T02, SSSA03, HS03, WT04, WT05}.

Understanding the properties of the physical driver of the acceleration of
the universe is of fundamental importance. The 
driver, commonly referred to as the dark energy, and its properties
are parameterized in quantities such as the 
potential and kinetic energy, and
the pressure, 
energy density, and equation of state.    
These quantities can be
expressed as combinations of the first and second derivative
of the coordinate distance, as discussed by \cite{dd04}.  

Here, we show
that the dark energy potential and kinetic energy may be 
determined as functions of redshift by appropriately combining
the first and second derivatives of the coordinate distance
to sources at different redshift.  This expansion of the 
method is described in 
section \ref{sec:PK}. The method is applied to an enlarged
data sample which includes  
the 71 new Legacy supernovae coordinate distances, presented
in section \ref{sec:newdata}, and the coordinate distances listed
in \cite{dd04}.  

The work presented here on the dark energy potential, $V(z)$, and
kinetic, $K(z)$, energy 
is complementary to the
work of \cite {SRSS00}, \cite{SVJ05} and \cite{SLP05}.
Saini et al. \cite{SRSS00} derive equations for $V(z)$ and $K(z)$ 
and use a fitting function for the luminosity distance
to obtain values and uncertainties for the parameters
of the fitting function; these are then used to obtain
$V(z)$, $K(z)$, and the equation of state parameter $w(z)$. 
Simon et al. \cite{SVJ05} consider the reconstruction of the 
dark energy potential based on an expansion of the 
dark energy potential in terms of Chebyshev polynomials,
while \cite{SLP05} consider the expansion of the quintessence potential
$V$ as a power series of the quintessence field $\phi$.  
\cite{SRSS00}, \cite{SVJ05} and \cite{SLP05} find that the results
are consistent with those expected if a cosmological constant
is driving the acceleration of the universe at the present epoch.  

\section{\label{sec:PK}The Potential and Kinetic Energy of 
the Dark Energy}

The potential and kinetic energy of the dark energy can be
determined as functions of redshift using 
measurements of the coordinate distance.  Measurements of the 
coordinate distance obtained using type Ia supernovae and type IIb
radio galaxies allow determinations of the first and second 
derivative of the coordinate distance with respect to redshift.
For detailed discussion of numerical methods used to derived these
quantities as well as extensive tests of the methods, see
\cite{dd03, dd04}.  
These derivatives may be combined to obtain $V(z)$ and $K(z)$, 
the potential and kinetic energy density of the dark energy,
which, for convenience, are expressed in units
of the critical density at the current epoch.

It is well known that $V = 0.5 (\rho-P)$ and 
$K = 0.5 (\rho+P)$, where $\rho$ and $P$ are the
energy density and pressure of the dark energy.  
In \cite{dd04} we show that 
both $\rho$ and $P$ may be written in terms of the first and
second derivatives of the coordinate distance (see Eqs. [6] and
[7] of \cite{dd04}).  Combining these, we find that 
\begin{equation}
\left({V \over \rho_{oc}}\right) =(y^\prime)^{-2}~[1+(1+z)(y^{\prime \prime})(y^\prime)^{-1}/3] - 0.5 \Omega_{0m} (1+z)^3~
\label{V}
\end{equation}
and 
\begin{equation}
\left({K \over \rho_{oc}}\right) = -(1/3)(1+z)(y^{\prime \prime})(y^\prime)^{-3}-0.5~\Omega_{0m}(1+z)^3~,
\label{K}
\end{equation}
where $\rho_{oc}$ is the critical density at the current epoch,
the dimensionless coordinate distance $y(z)\equiv H_0(a_0r)$, 
$H_0$ is Hubble's constant, $(a_0r)$ is the coordinate distance
to a source at redshift z, 
$y^\prime \equiv (dy/dz)$, and $y^{\prime \prime} \equiv (d^2y/dz^2)$.

In obtaining Eqs. [\ref{V}] and [\ref{K}], 
it has been assumed that: the universe
is spatially flat; the kinematics of the universe are accurately
described by general relativity; and two components, the dark
energy and non-relativisitc matter, are sufficient to 
account for the kinematics of the universe out to redshift of about 2
(see the discussion in \cite{dd04}).  

Thus, the redshift behavior of the 
potential and kinetic energy densities of the dark energy 
can be constructed using the first and second derivatives of the 
dimensionless coordinate distance $y(z)$, for the data set described below.  

\section{\label{sec:newdata} The Data Set}

The core of our data set remains the sample of 157 ``Gold'' supernovae
of \cite{R04} and the 20 radio galaxids of \cite{GDW00}, both of which  are
tabulated in \cite{dd04}.  This data set is supplemented by 71 new
supernvae from the Supernova Legacy Survey of \cite{A05},
which allows the determination of 
dimensionless
coordinate distances to 71 additional type Ia supernovae.  These
are listed in Table I, and were obtained using the values and uncertainties
of $\mu_B$ listed in Table 9 of \cite{A05}, given the value
of $H_0 = 70$ km/s/Mpc adopted by \cite{A05}.   Two uncertainties
of $y$ are listed: $\sigma(y)$ is obtained from the uncertainty in
$\mu_B$ listed in Table 9 of \cite{A05}, while the total uncertainty
of y (used throughout this paper) $\sigma_T(y)$ reflects the  
total uncertainty of $\mu_B$, which is obtained by adding in quadrature 
the uncertainty listed
in Table 9 to the intrinsic dispersion of 0.13 identified by \cite{A05}.  
These are listed in Table \ref{table1}.   

\begin{figure}
\includegraphics[width=90mm]{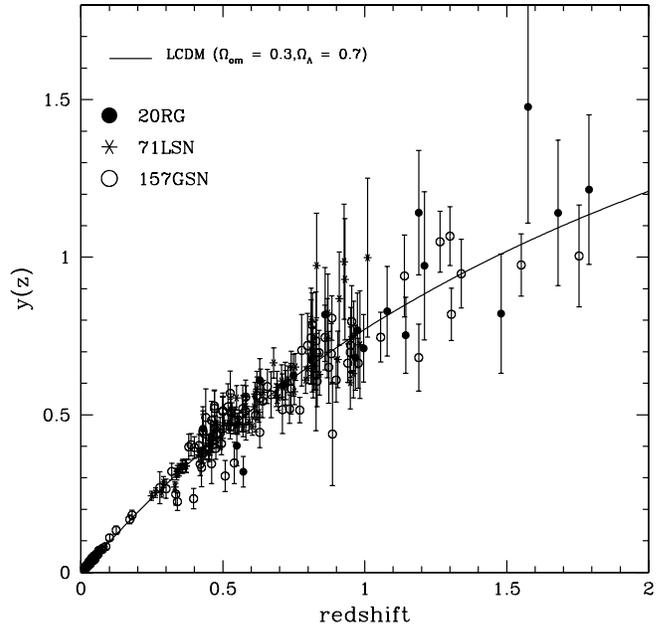}
\caption{\label{Y248}Coordinate distances to the 71 Legacy and 
157 Gold supernovae and 20 radio galaxies. }
\end{figure}

The total sample of 248 sources, including 228 supernovae
and 20 radio galaxies, is shown in Fig. \ref{Y248}.   We note that there
are no systematic differences seen among the three groups of measurements
in the redshift ranges of their overlaps.

\begin{table}
\caption{\label{table1} Coordinate distances to the 71 Legacy Supernovae
and their uncertainties. }
\begin{ruledtabular}
\begin{tabular}{cccccccc}
name	&	z	&	y	&	$\sigma$ (y)	&	$\sigma_T$ (y)	\\
\hline
SNLS-03D1au	&	0.504	&	0.475	&	0.0085	&	0.0297	\\
SNLS-03D1aw	&	0.582	&	0.556	&	0.0138	&	0.0360	\\
SNLS-03D1ax	&	0.496	&	0.426	&	0.0074	&	0.0265	\\
SNLS-03D1bp	&	0.346	&	0.325	&	0.0031	&	0.0197	\\
SNLS-03D1cm	&	0.870	&	0.822	&	0.1140	&	0.1242	\\
SNLS-03D1co	&	0.679	&	0.665	&	0.0269	&	0.0480	\\
SNLS-03D1ew	&	0.868	&	0.743	&	0.1177	&	0.1258	\\
SNLS-03D1fc	&	0.331	&	0.271	&	0.0016	&	0.0163	\\
SNLS-03D1fl	&	0.688	&	0.562	&	0.0127	&	0.0360	\\
SNLS-03D1fq	&	0.800	&	0.647	&	0.0268	&	0.0471	\\
SNLS-03D1gt	&	0.548	&	0.554	&	0.0204	&	0.0389	\\
SNLS-03D3af	&	0.532	&	0.502	&	0.0192	&	0.0357	\\
SNLS-03D3aw	&	0.449	&	0.380	&	0.0077	&	0.0240	\\
SNLS-03D3ay	&	0.371	&	0.338	&	0.0047	&	0.0207	\\
SNLS-03D3ba	&	0.291	&	0.286	&	0.0044	&	0.0177	\\
SNLS-03D3bh	&	0.249	&	0.243	&	0.0022	&	0.0147	\\
SNLS-03D3cc	&	0.463	&	0.417	&	0.0065	&	0.0258	\\
SNLS-03D3cd	&	0.461	&	0.407	&	0.0109	&	0.0267	\\
SNLS-03D4ag	&	0.285	&	0.254	&	0.0018	&	0.0153	\\
SNLS-03D4at	&	0.633	&	0.605	&	0.0178	&	0.0404	\\
SNLS-03D4cn	&	0.818	&	0.653	&	0.0914	&	0.0994	\\
SNLS-03D4cx	&	0.949	&	0.602	&	0.0754	&	0.0836	\\
SNLS-03D4cy	&	0.927	&	0.986	&	0.1725	&	0.1823	\\
SNLS-03D4cz	&	0.695	&	0.554	&	0.0219	&	0.0398	\\
SNLS-03D4dh	&	0.627	&	0.508	&	0.0082	&	0.0315	\\
SNLS-03D4di	&	0.905	&	0.676	&	0.0803	&	0.0899	\\
SNLS-03D4dy	&	0.604	&	0.463	&	0.0062	&	0.0284	\\
SNLS-03D4fd	&	0.791	&	0.610	&	0.0214	&	0.0423	\\
SNLS-03D4gf	&	0.581	&	0.526	&	0.0114	&	0.0335	\\
SNLS-03D4gg	&	0.592	&	0.477	&	0.0198	&	0.0347	\\
SNLS-03D4gl	&	0.571	&	0.462	&	0.0149	&	0.0314	\\
SNLS-04D1ag	&	0.557	&	0.476	&	0.0064	&	0.0292	\\
SNLS-04D1aj	&	0.721	&	0.594	&	0.0290	&	0.0459	\\
SNLS-04D1ak	&	0.526	&	0.517	&	0.0131	&	0.0336	\\
SNLS-04D2cf	&	0.369	&	0.338	&	0.0025	&	0.0204	\\
SNLS-04D2fp	&	0.415	&	0.373	&	0.0046	&	0.0228	\\
SNLS-04D2fs	&	0.357	&	0.334	&	0.0028	&	0.0202	\\
SNLS-04D2gb	&	0.430	&	0.370	&	0.0065	&	0.0231	\\
SNLS-04D2gc	&	0.521	&	0.472	&	0.0117	&	0.0306	\\
SNLS-04D2gp	&	0.707	&	0.607	&	0.0361	&	0.0512	\\
SNLS-04D2iu	&	0.691	&	0.587	&	0.0368	&	0.0509	\\
SNLS-04D2ja	&	0.741	&	0.649	&	0.0350	&	0.0523	\\
SNLS-04D3co	&	0.620	&	0.581	&	0.0161	&	0.0383	\\
SNLS-04D3cp	&	0.830	&	0.973	&	0.1556	&	0.1661	\\
SNLS-04D3cy	&	0.643	&	0.571	&	0.0155	&	0.0376	\\
SNLS-04D3dd	&	1.010	&	0.999	&	0.2451	&	0.2523	\\
SNLS-04D3df	&	0.470	&	0.451	&	0.0066	&	0.0278	\\
SNLS-04D3do	&	0.610	&	0.525	&	0.0094	&	0.0328	\\
SNLS-04D3ez	&	0.263	&	0.253	&	0.0015	&	0.0152	\\
SNLS-04D3fk	&	0.358	&	0.339	&	0.0020	&	0.0204	\\
SNLS-04D3fq	&	0.730	&	0.613	&	0.0212	&	0.0424	\\
SNLS-04D3gt	&	0.451	&	0.411	&	0.0057	&	0.0253	\\
SNLS-04D3gx	&	0.910	&	0.868	&	0.1384	&	0.1478	\\
SNLS-04D3hn	&	0.552	&	0.467	&	0.0075	&	0.0290	\\
SNLS-04D3is	&	0.710	&	0.589	&	0.0209	&	0.0410	\\
SNLS-04D3ki	&	0.930	&	0.930	&	0.1841	&	0.1924	\\
SNLS-04D3kr	&	0.337	&	0.312	&	0.0014	&	0.0187	\\
SNLS-04D3ks	&	0.752	&	0.573	&	0.0238	&	0.0418	\\
SNLS-04D3lp	&	0.983	&	0.723	&	0.1650	&	0.1706	\\
\end{tabular}
\end{ruledtabular}
\end{table}

\begin{table}
\caption{ Table 1 \it{Continued} }
\begin{ruledtabular}
\begin{tabular}{cccccccc}
name	&	z	&	y	&	$\sigma$ (y)	&	$\sigma_T$ (y)	\\
\hline
SNLS-04D3lu	&	0.822	&	0.655	&	0.0658	&	0.0766	\\
SNLS-04D3ml	&	0.950	&	0.739	&	0.0912	&	0.1014	\\
SNLS-04D3nc	&	0.817	&	0.690	&	0.0807	&	0.0907	\\
SNLS-04D3nh	&	0.340	&	0.320	&	0.0018	&	0.0193	\\
SNLS-04D3nr	&	0.960	&	0.631	&	0.0680	&	0.0778	\\
SNLS-04D3ny	&	0.810	&	0.706	&	0.0978	&	0.1065	\\
SNLS-04D3oe	&	0.756	&	0.652	&	0.0174	&	0.0427	\\
SNLS-04D4an	&	0.613	&	0.566	&	0.0159	&	0.0374	\\
SNLS-04D4bk	&	0.840	&	0.628	&	0.0535	&	0.0654	\\
SNLS-04D4bq	&	0.550	&	0.473	&	0.0122	&	0.0308	\\
SNLS-04D4dm	&	0.811	&	0.794	&	0.0966	&	0.1077	\\
SNLS-04D4dw	&	0.961	&	0.751	&	0.1003	&	0.1099	\\
\end{tabular}
\end{ruledtabular}
\end{table}

\section{\label{sec:Fullsample}Results Obtained with the Full Sample of 
248 Sources}

\begin{figure}
\includegraphics[width=90mm]{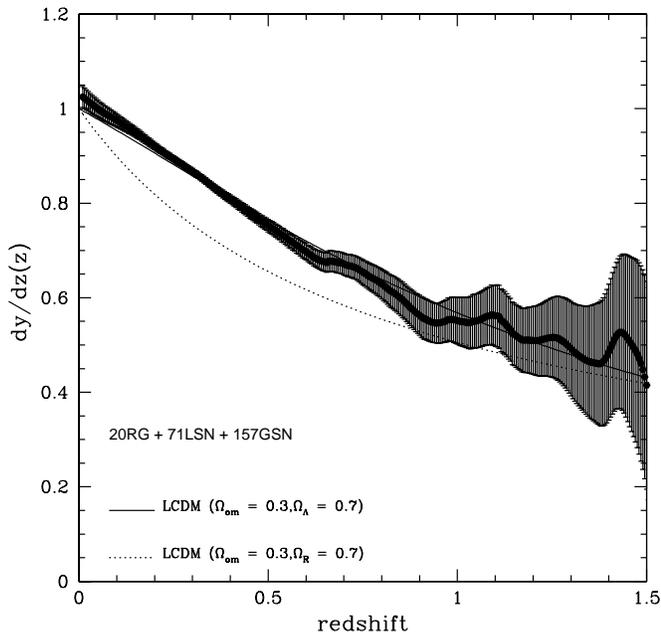} 
\caption{\label{dydz248snrg} The first derivative of the coordinate 
distance
with respect to redshift. 
The values for the standard LCDM model with 
$\Omega_{\Lambda} = 0.7$ and $\Omega_{0m}=0.3$ are shown as the
solid line in this and all subsequent plots.
The values for the curvature-dominated model described in the
text are shown with the dotted line.
The zero redshift
value we measure is $y ^\prime _0 = 1.025 \pm 0.022$. The predicted value is 1.000.} 
\end{figure}

\begin{figure}
\includegraphics[width=90mm]{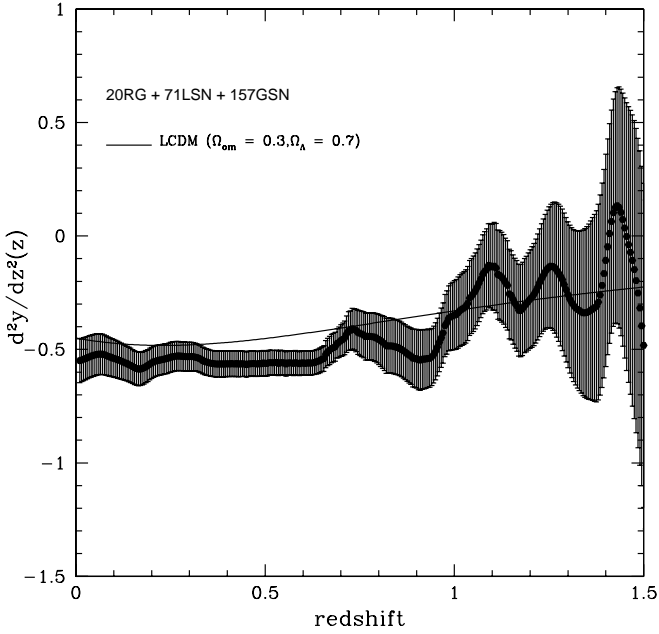}
\caption{\label{d2ydz2snrg248}The second derivative of the 
coordinate distance with 
respect to redshift. The measured zero redshift
value is $y ^{\prime \prime} _0 = -0.55 \pm 0.10$; the value predicted
in a standard LCDM model is $-0.45$.
}
\end{figure}

Values of $y^{\prime}$
and $y^{\prime \prime}$ obtained with the full sample are shown in Figs.
[\ref{dydz248snrg}], and [\ref{d2ydz2snrg248}].
As in \cite{dd03,dd04}, a window function with $\Delta z$ of 0.6 
is used in this paper.    
Note that the first and second derivatives are obtained directly
from the measured values of $y$; no assumptions have been adopted
regarding a theory of gravity, or the form of the expansion or
acceleration rate of the universe as a function of redshift.  
For comparison, the values of parameters predicted 
in a standard 
lambda cold dark matter (LCDM) cosmological model, where the kinematics 
of the universe are described by general relativity and
the primary components of the universe at present 
are a normalized cosmological constant 
$\Omega_{\Lambda} = 0.7$ and non-relativistic
matter $\Omega_{0m}=0.3$, are shown
on each figure. In addition, the curve expected in a universe 
with space curvature $k$ is shown in Fig. [\ref{dydz248snrg}], where
values of $\Omega_{0m} = 0.3$, $\Omega_{\Lambda} = 0$, 
and $\Omega_R = -k/(a_0H_0)^2 = 0.7$ have been assumed.
Clearly, this curvature dominated model is inconsistent with the results obtained here. 

\begin{figure}
\includegraphics[width=90mm]{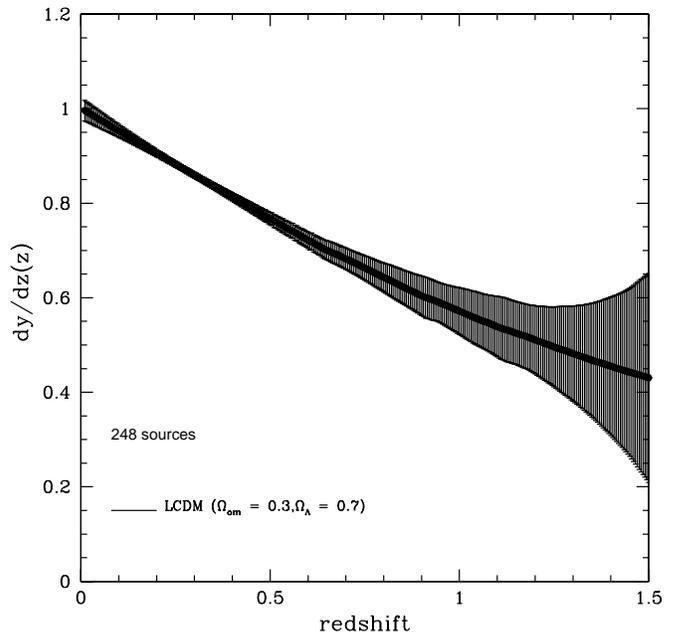} 
\caption{\label{dydz248compare} As in Fig. \ref{dydz248snrg}, 
but for the mock data set with the same redshift and error distribution
as the actual data set, with the assumed cosmology with
$\Omega_{\Lambda} = 0.7$ and $\Omega_{0m}=0.3$.
The fit results for this mock data set are in an excellent agreement
with the analytical prediction for the assumed cosmology, giving us
a confidence that the numerical method we use is unbiased and is
capable of recovering accurately the underlying cosmology.
 } 
\end{figure}

\begin{figure}
\includegraphics[width=90mm]{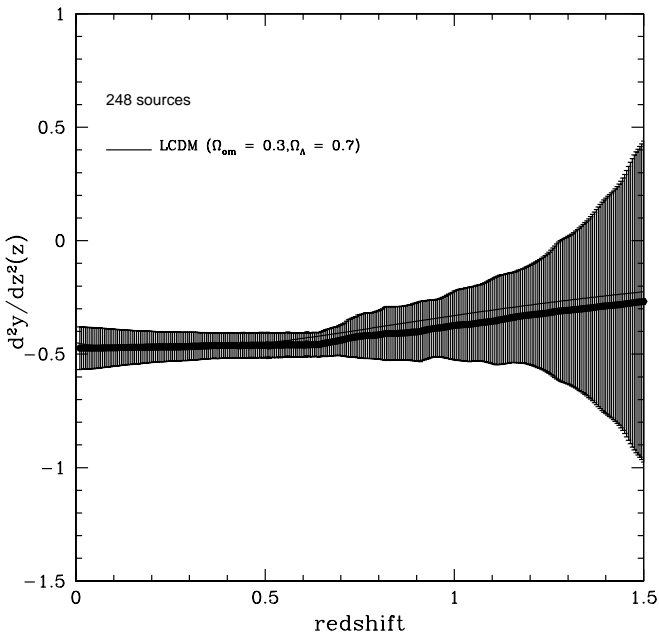}
\caption{\label{d2ydz2compare248} 
As in Fig. \ref{d2ydz2snrg248},
but for the same mock data set.
Again, the correct assumed cosmology is recovered with a
negligible bias.
}
\end{figure}

To compare our results with those expected in a LCDM model a mock
data set of 248 sources was constructed in which the mock 
sources have the same 
redshift distribution and fractional error per point as the empirical
data but have values of $y$ predicted in a LCDM model with 
$\Omega_{\Lambda} = 0.7$ and $\Omega_{0m}=0.3$.  This mock data
set was passed through the same numerical programs as the 
true data, and the results obtained for $y^{\prime}$ and $y^{\prime \prime}$
for both the true and mock data sets
are shown in Figs. [\ref{dydz248compare}] and [\ref{d2ydz2compare248}].  
It is quite clear that no bias is introduced by the numerical differentiation;
the central value of $y^{\prime}$ and $y^{\prime \prime}$
output by the numerical differentiation 
lie on those predicted in the assumed LCDM model, and the magnitude of
the error bars at a given redshift have the same magnitude as 
those obtained from the data.


\begin{figure}
\includegraphics[width=90mm]{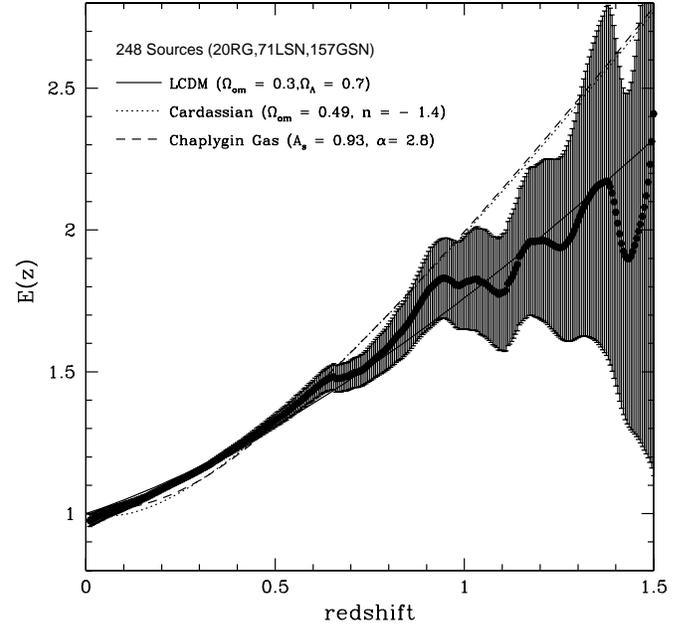} 
\caption{\label{E248} The dimensionless expansion parameter
$E(z)$.  The measured zero redshift
value is $E_0 = 0.98 \pm 0.02$. The predicted value is 1.00.
The solid line shows the prediction of the standard
$\Omega_{\Lambda} = 0.7$ and $\Omega_{0m}=0.3$ model,
which is in an excellent agreement with the data.
The dotted line represents predictions of the Cardassian model,
and the dashed line represents the Chaplygin gas model, obtained
assuming that the universe is spatially flat, 
both of which seem to fit the data systematically less well.
} 
\end{figure}

\begin{figure}
\includegraphics[width=90mm]{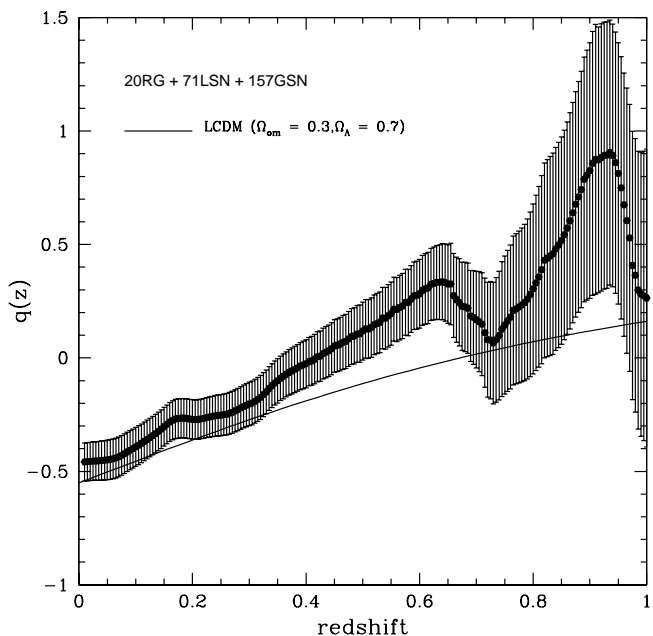} 
\caption{\label{Q248} The deceleration parameter $q(z)$. 
The zero redshift
value is $q_0 = -0.46 \pm 0.08$. The predicted value in a standard
LCDM model is $-0.55$.
Our fits are systematically too high by about 1$\sigma$ relative
to the standard model.
} 
\end{figure}

Values of the dimensionless expansion rate $E(z)$,
and the deceleration parameter $q(z)$ are shown 
in Figs. [\ref{E248}] and [\ref{Q248}], obtained using
equations [5] and [6] of \cite{dd03}.  The only assumptions
that must be adopted to construct $E$ and $q$ from 
$y ^{\prime}$ and $y ^{\prime \prime}$ are that the universe
is homogeneous and isotropic on large scales, and
is spatially flat (see \cite{dd03}).  Again, the curve expected in a 
LCDM model with $\Omega_{\Lambda}=0.7$ and $\Omega_{0m}=0.3$
is shown in each figure. In addition, the curves predicted
in two modified gravity models is a spatially flat universe
are included in Fig. [\ref{E248}].
These are obtained using the best fit model parameters obtained
by \cite{BBSS05} for the Cardassian model of \cite{FL02}
and the generalized Chaplygin gas model of \cite{BBS02} based
on the model of \cite{KMP01}.  Clearly, the LCDM model provides
a better description of the data than do either of the modified
gravity models.  

\begin{figure}
\includegraphics[width=90mm]{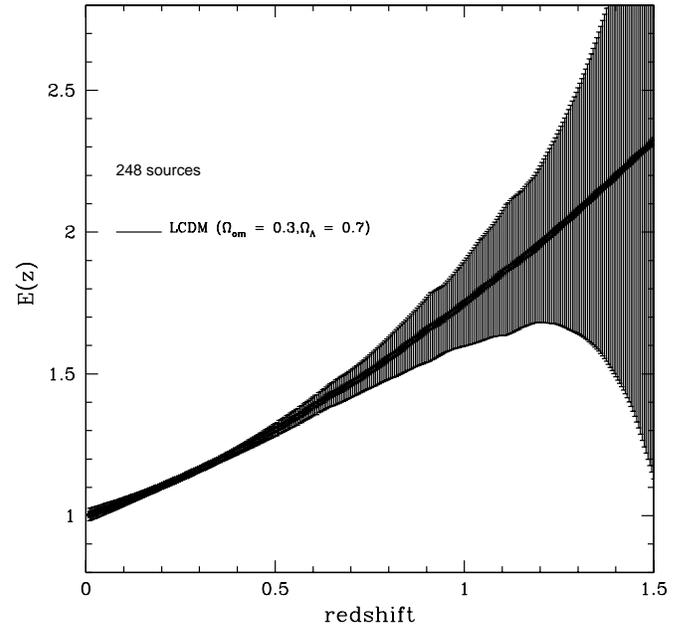} 
\caption{\label{E248compare} Derived values of $E(z)$
for the mock data set.  Again, the correct cosmology is
recovered accurately by the method.} 
\end{figure}

\begin{figure}
\includegraphics[width=90mm]{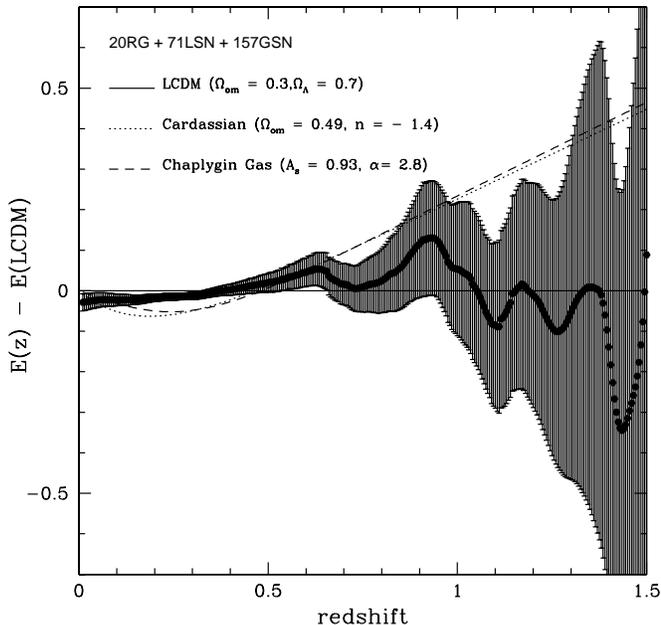} 
\caption{\label{E248diff} Values of E(z) relative to the 
LCDM model with 
$\Omega_{\Lambda} = 0.7$ and $\Omega_{0m}=0.3$,
which has been subtracted from the data and the 
models. This shows how well the LCDM model describes
E(z) and that the  
Cardassian model and generalized Chaplygin gas model in a spatially
flat universe seem to fit the data systematically less well.
} 
\end{figure}

\begin{figure}
\includegraphics[width=90mm]{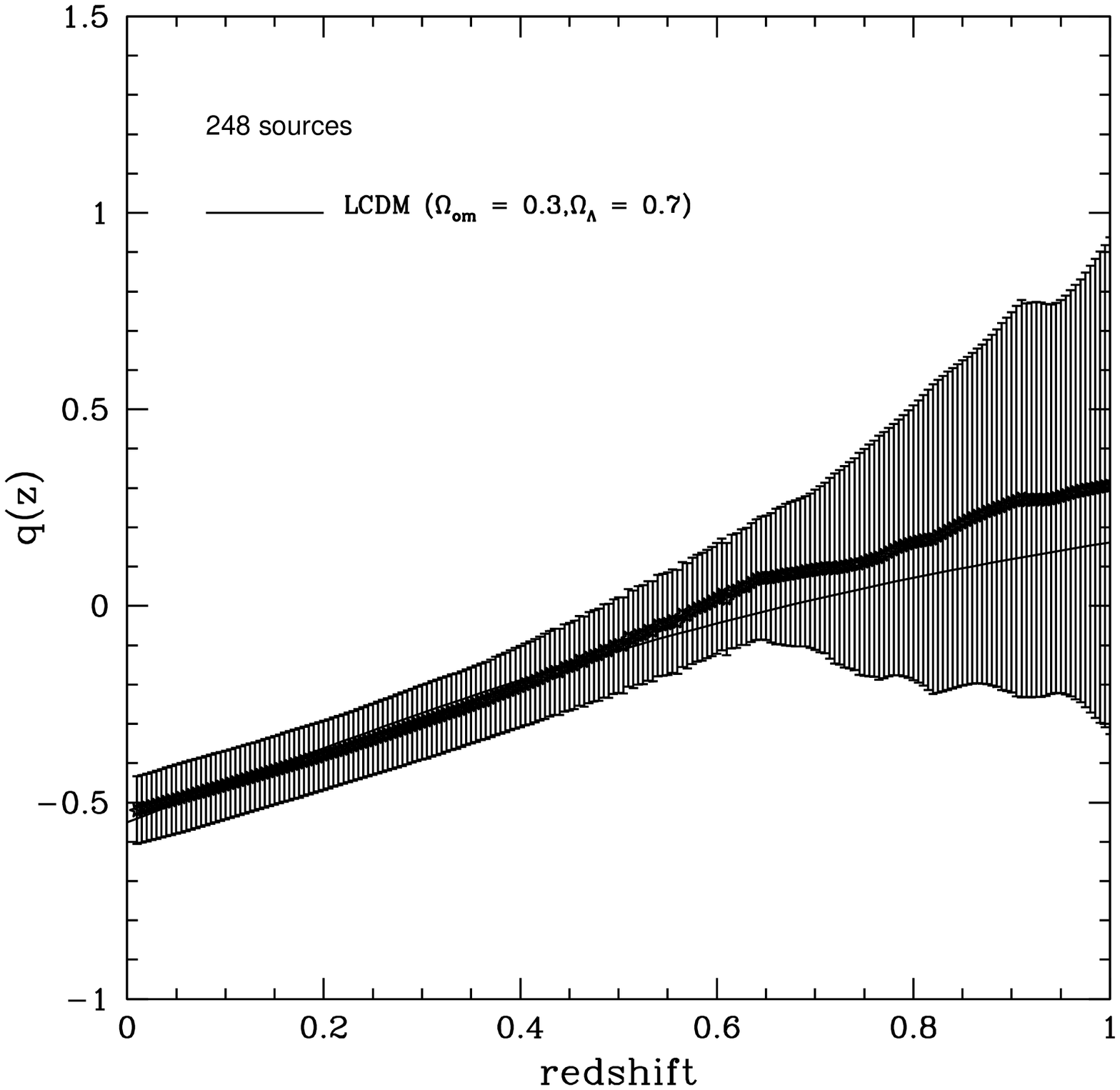} 
\caption{\label{Q248compare} 
Derived values of the $q(z)$, but for the mock data set.
Again, the underlying cosmology is recovered, except for a
slight bias at the high redshift end.
} 
\end{figure}

Fig. [\ref{E248diff}] shows the difference between 
values of $E(z)$ and those expected in a LCDM model with 
$\Omega_{\Lambda} = 0.7$ and $\Omega_{0m}=0.3$.  
This further illustrates how well General Relativity 
with a cosmological constant describes the data over the
very large length scale of greater than about 10 billion light years. 
Recall that $E(z)$ is derived from the data without having
to specify a theory of gravity.   

The results shown for $q(z)$ allow
a determination of the redshift at which  
the universe transitions from an accelerating phase
to a decelerating phase; we find this to be  
at a redshift of $z_T = 0.42 \pm {}^{0.08}_{0.06}$, consistent
with the values quoted by \cite{dd03,dd04} and \cite{R04}. 
The upper bound on this transition redshift is uncertain
because of the fluctuations in $q(z)$ which are due to 
sparse sampling at high redshift.  In this plot, and in the ones that
follow, we do not consider these fluctuations at higher redshifts
to be statistically significant, as they are commensurate
with our derived 1-$\sigma$ error bars. 

A comparison with these results and those expected in a LCDM
model is shown in Figs. [\ref{E248compare}] and [\ref{Q248compare}],
where the predicted values of $E(z)$ and $q(z)$ are obtained using the
mock data set described above.  Again, we see that no bias has been 
introduced by the numerical differentiation technique, and 
the data are in very good agreement with the standard LCDM model. 

\begin{figure}
\includegraphics[width=90mm]{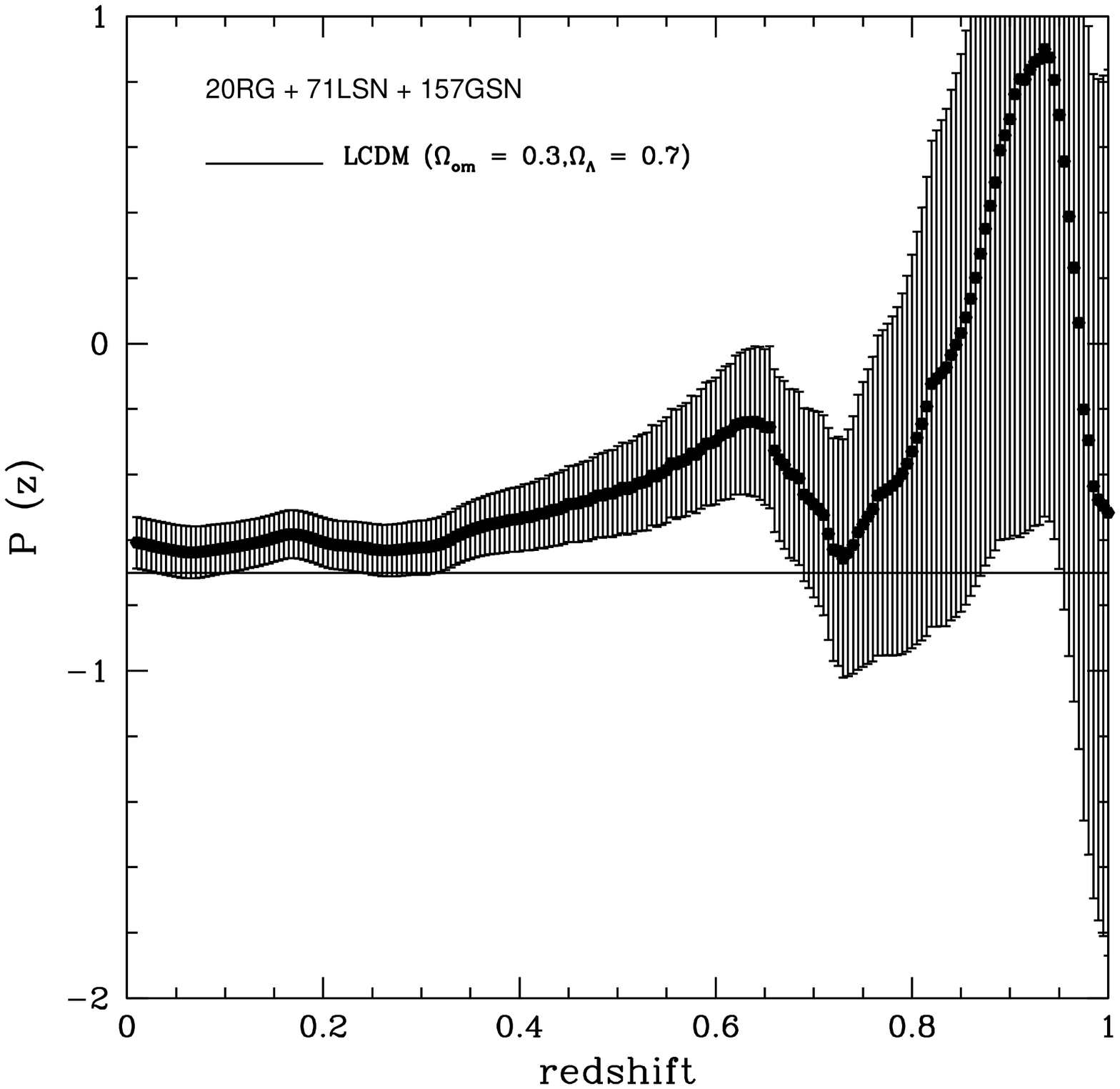} 
\caption{\label{p248} The pressure of the dark energy in units of
the critical density today. 
The zero redshift
value of $p_0 = -0.61 \pm 0.08$.}  
\end{figure}

To obtain the pressure, a theory of gravity must be specified,
and general relativity has been assumed here (see eqs. [5]
and [6] of \cite{dd04}).  Note that if the dark energy is
a cosmological constant, the zero redshift value of $p = P/\rho_{oc}$ is
a measure of $\Omega_{\Lambda}$.  The pressure 
of the dark energy, in units of the critical density today, 
obtained with 
the current data set is shown in Fig. [\ref{p248}].  The zero
redshift value suggests $\Omega_{\Lambda}=0.61 \pm 0.08$,
in excellent agreement with the values commonly derived using
more traditional approaches.

\begin{figure}
\includegraphics[width=90mm]{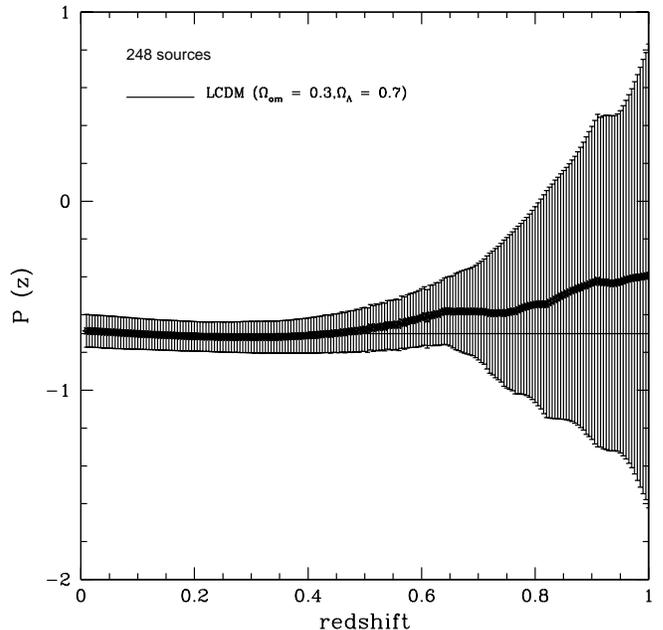} 
\caption{\label{p248compare} Values of the $P(z)$, but for the
mock data set.  The assumed cosmology is recovered accurately.}  
\end{figure}

Values of $p(z)$ expected 
in a LCDM model are shown in Fig. [\ref{p248compare}], where the 
predicted values of $p(z)$ are obtained using the mock data
set described above.  Again, there is no bias introduced by
the numerical differentiation technique, and the output is 
in good agreement with predictions in a LCDM model. 

\begin{figure}
\includegraphics[width=90mm]{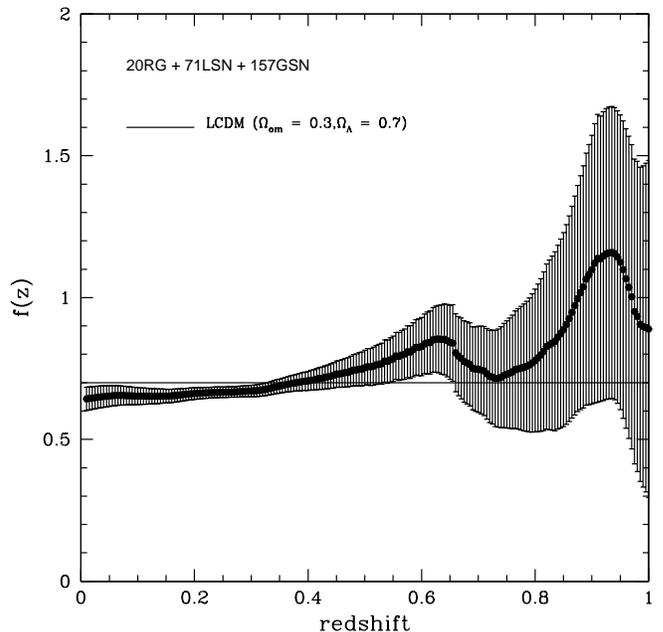} 
\caption{\label{f248} The energy density of the dark energy
in units of the critical density today, $f(z)$. 
The zero redshift
value is $f_0 = 0.64 \pm 0.04$, again in an excellent
agreement with the more traditional determinations of
$\Omega_{\Lambda}$.}  
\end{figure}

\begin{figure}
\includegraphics[width=90mm]{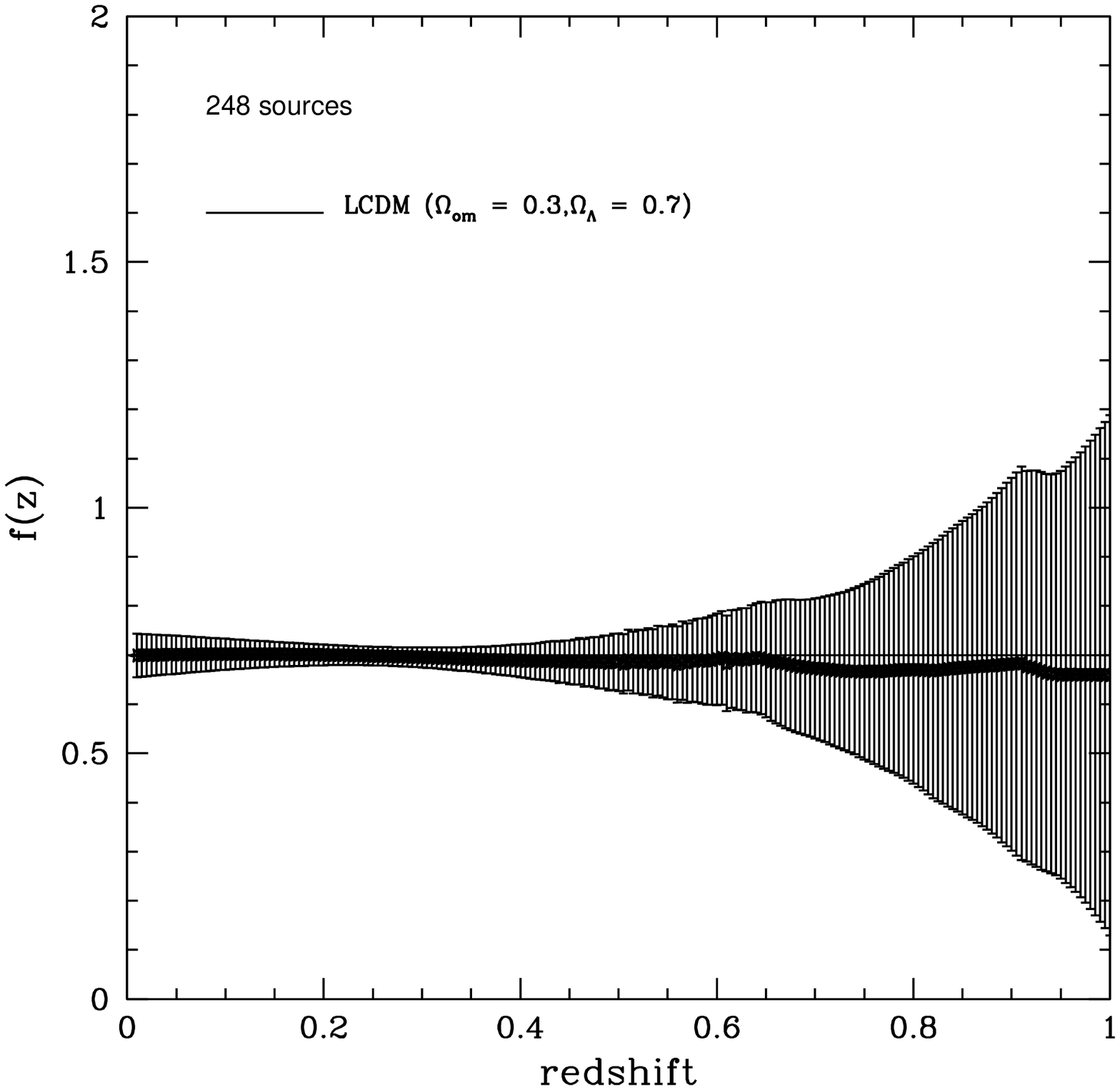} 
\caption{\label{f248compare} The values of $f(z)$ for the 
mock data set.  The assumed value of 0.7 is recovered
accurately.}
\end{figure}

The energy density $f(z) = \rho/\rho_{oc}$ of the dark energy in
units of the critical density today
can be obtained once
a value for $\Omega_{0m}$ has been adopted.  The value
of $f(z)$ obtained assuming 
$\Omega_{0m}=0.3$ is shown in Fig. [\ref{f248}]. These results
are compared with predictions in a LCDM model obtained with  
the mock data set 
in Fig. [\ref{f248compare}].
There is good agreement between the data and predictions in a
LCDM model.  

\begin{figure}
\includegraphics[width=90mm]{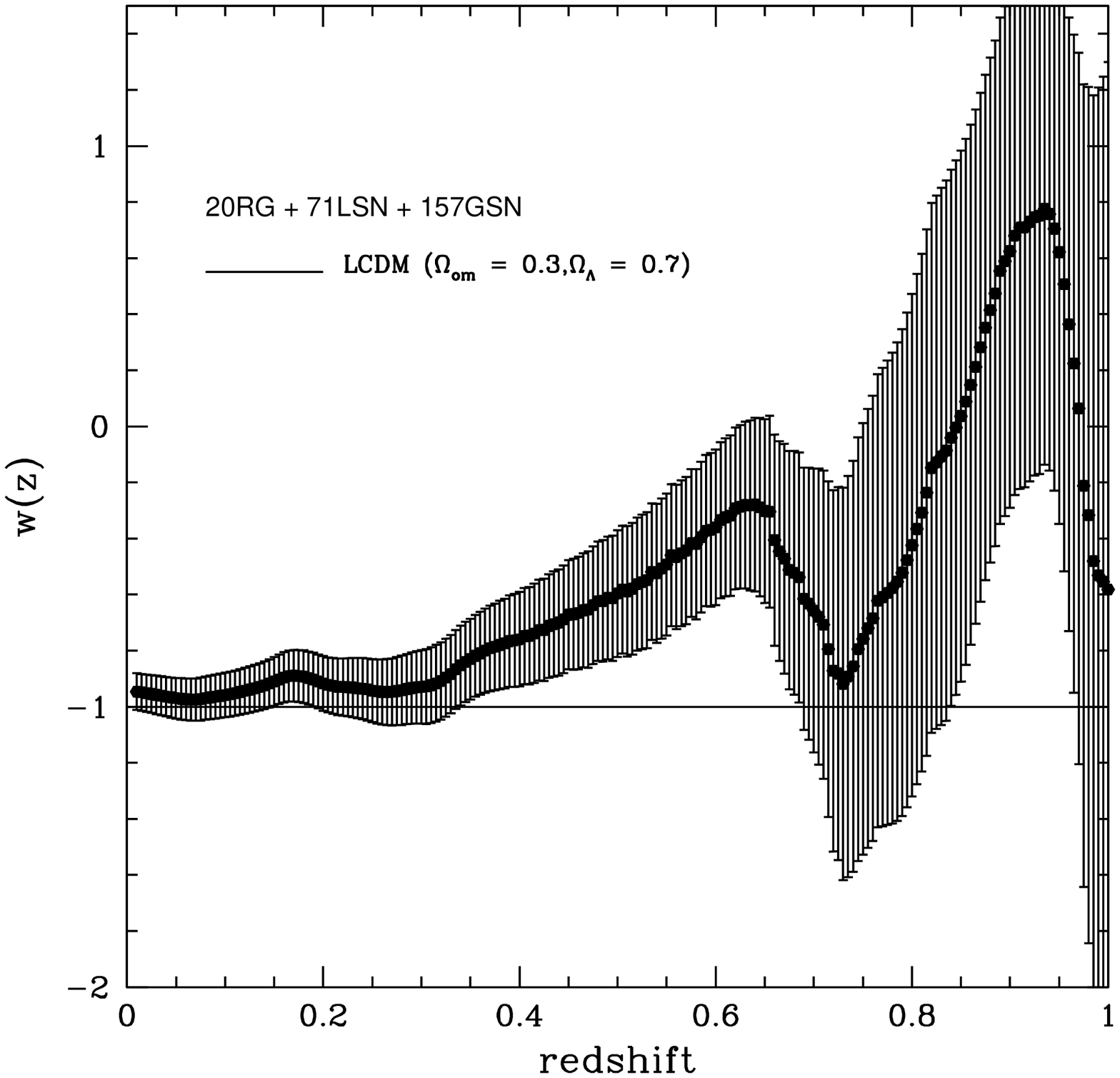} 
\caption{\label{w248} The dark energy equation of state parameter 
$w(z)$.  The zero redshift
value is $w_0 = -0.95 \pm 0.07$, whereas $w = -1$ is the theoretical
value for the cosmological constant model.}  
\end{figure}

\begin{figure}
\includegraphics[width=90mm]{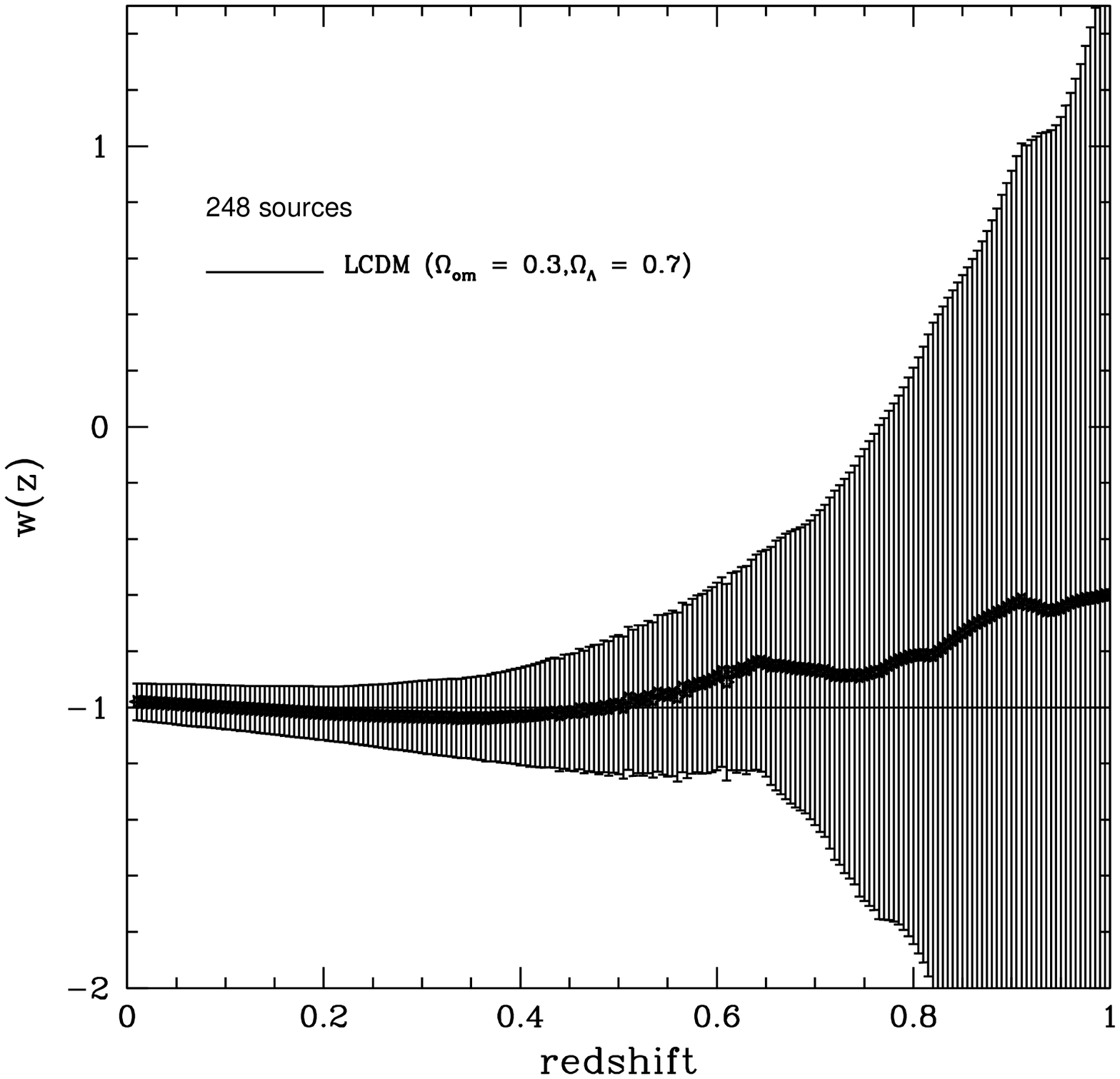} 
\caption{\label{w248compare} The values of $w(z)$ for the mock data
set.  The assumed value of $w = -1$ is recovered accurately, with
only a modest bias at the high redshift end.}  
\end{figure}

The equation of state of the dark energy, $w = \rho/P$ is
shown in Fig. [\ref{w248}].  Results obtained with the 
mock data set obtained in a LCDM model 
are shown in Fig. [\ref{w248compare}].
Figs. [\ref{E248}, \ref{Q248},
\ref{p248}, \ref{f248}] and [\ref{w248}] provide 
an update on the results presented by \cite{dd04}; 
the dependence of each quantity on the first and second 
derivatives of the dimensionless coordinate distance $y$ with 
respect to redshift $z$ is given in \cite{dd04}.

\begin{figure}
\includegraphics[width=90mm]{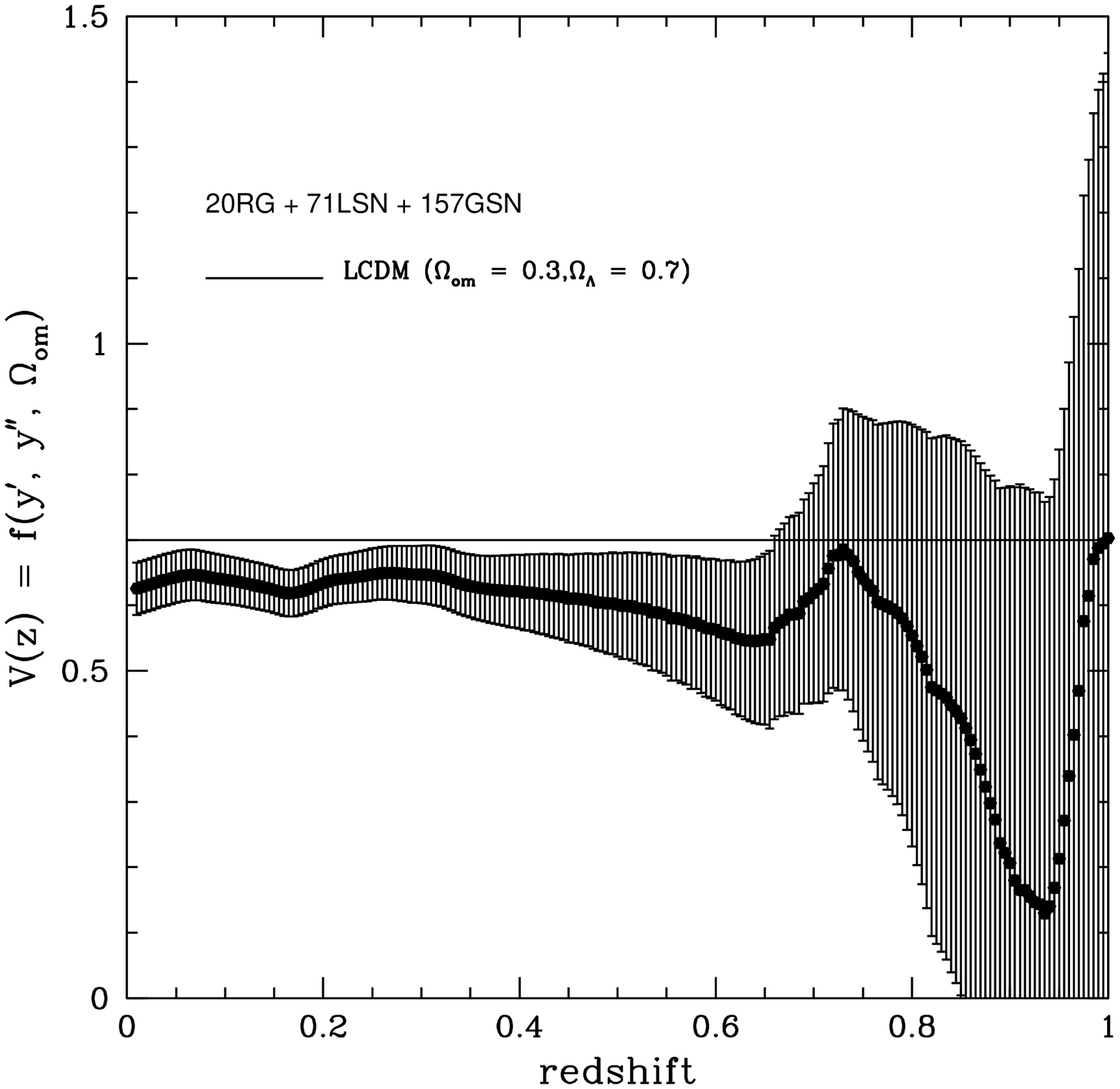} 
\caption{\label{V248} The potential energy of the dark energy 
$V(z)$ in units of the critical density today. The zero
redshift value is $V_0 = 0.63 \pm 0.05$, whereas the expected
value for the standard LCDM model is 0.7.} 
\end{figure}

\begin{figure}
\includegraphics[width=90mm]{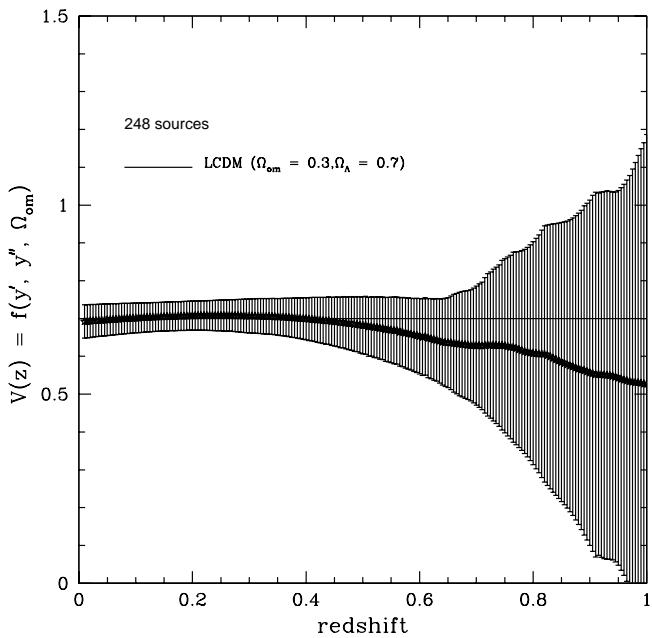} 
\caption{\label{V248compare} The values of 
$V(z)$ for the mock data set.  The assumed value of 0.7 is
recovered accurately, except for a small bias at the high
redsihfts.} 
\end{figure}

\begin{figure}
\includegraphics[width=90mm]{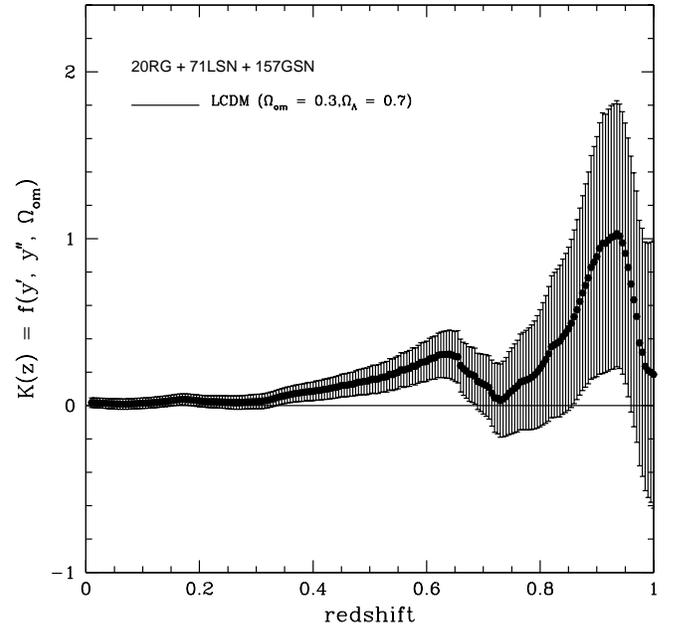} 
\caption{\label{K248} The kinetic energy  of the dark
energy K(z) in units of the critical density today. 
The zero redshift value is $K_0 = 0.02 \pm 0.03$, whereas
for the standard LCDM model the expected value is 0.} 
\end{figure}

\begin{figure}
\includegraphics[width=90mm]{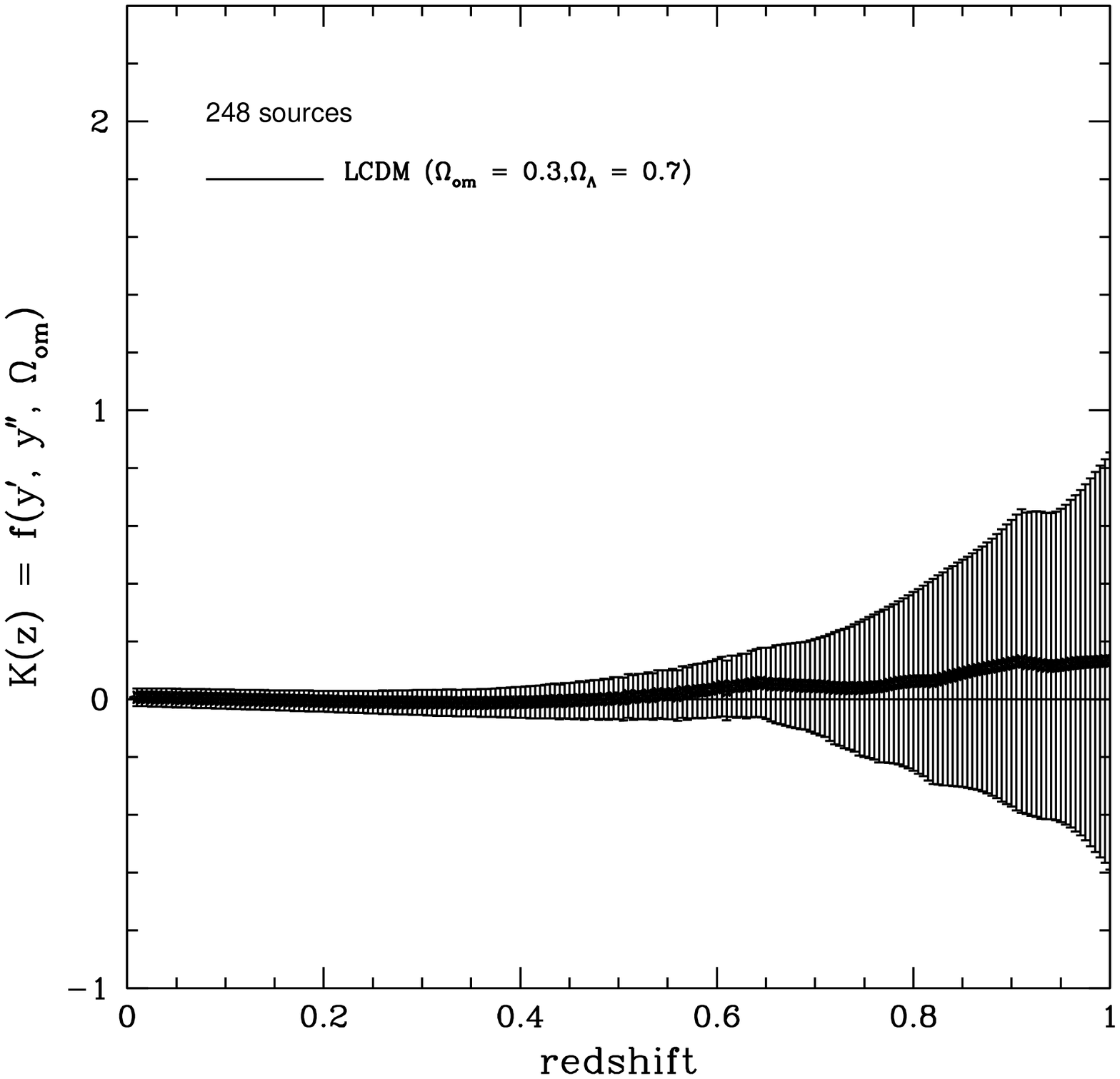} 
\caption{\label{K248compare} The values of $K(z)$ for the
mock data set. Again, the correct value for the assumed
cosmology, $K = 0$, is recovered accurately.} 
\end{figure}

The values of $V$ and $K$ obtained with the full sample
of 248 sources are shown in Figs. [\ref{V248}] and [\ref{K248}].
The results are consistent with a dark energy potential
energy  that is constant from a redshift of zero
to a redshift of about 0.8, as expected if the
dark energy is a cosmological constant.   
The zero redshift normalization
depends on $y^\prime (z=0)$ and 
$y^{\prime \prime} (z=0)$ (see Eq. [\ref{V}]).  
The value of $y^{\prime \prime}$ at z=0 is 
$-0.55 \pm 0.1$, while that of $y^\prime$ is 
$1.03 \pm 0.02$; these cause the value of V(z=0) to be low
relative to a LCDM model with $\Omega_{0m}=0.3$.
When the number of low-redshift type Ia supernovae is
substantially increased, we expect to be able to determine
whether the normalization of $V$ differs from that expected
in a standard LCDM model.   These are compared with predictions
in a LCDM model using the mock data set,
as shown in Figs. [\ref{V248compare}] and [\ref{K248compare}];  a 
LCDM model provides a good description of the data. 


We do not understand completely
the small systematic difference between
our evaluations of some of the observed trends from the standard
LCDM model: our values being systematically too high for
the $q(z)$, $P(z)$, and $w(z)$, and too low for $V(z)$, by about
1-$\sigma$ throughout.  We are inclined to interpret this
as the inherent limitation of the current data set, rather than
as a real physical effect.

\section{\label{sec:level1}Summary and Conclusions}

The redshift behavior of the 
potential and kinetic energy  of the dark energy
can be determined directly from observed quantities
without having to assume an a priori functional form
for these quantities. They depend upon the first and
second derivatives of the coordinate distance with 
respect to redshift, and on the zero redshift value
of the mean mass density in non-relativistic matter.
These quantities, $y^\prime$, $y^{\prime \prime}$, $V(z)$, and
$K(z)$ were obtained using a sample of 248 sources including
228 type Ia supernovae and 20 type IIb radio galaxies. 
In all cases, the results are consistent with 
predictions of a LCDM model out to a redshift of about one.

The data may also be used to determine the deceleration
parameter $q(z)$ and the dimensionless expansion parameter $E(z)$
directly from $y^\prime$ and $y^{\prime \prime}$; these
determinations of $q$ and $E$ assume only that the universe
is homogeneous and isotropic on large scales, and is
spatially flat, and are independent of any assumptions
regarding a theory of gravity or the properties of the dark energy,
as discussed in detail by \cite{dd03,dd04}. We find that 
the universe transitions from an accelerating phase
to a decelerating phase at a redshift of $z_T = 0.42 \pm {}^{0.08}_{0.06}$.
The fact that $y, y^{\prime},$ and $y^{\prime \prime}$ match
predictions in a LCDM model to about one sigma or better to a
redshift of about 1.5 means that General Relativity with
a cosmological constant provides an accurate description of
the data on scales of about 10 billion light years. 
The correct explanation of the dark energy must be able to 
account for the observed values of 
$y(z)$, $y^\prime(z)$ and $y^{\prime \prime}(z)$, or, equivalently,
$y(z)$, $E(z)$, and $q(z)$.  

By adopting general relativity as the correct theory of gravity,
the pressure of the dark energy can be obtained as a function
of redshift. As shown by \cite{dd04}, this depends only upon
$y ^{\prime}$ and $y ^{\prime \prime}$; it is independent of 
assumptions regarding the properties of the dark energy
and independent of $\Omega_{0m}$.  If the universe is 
dominated by a cosmological constant at the present epoch,
the zero redshift value
of $p$ yields a new method of determining $\Omega_{\Lambda}$, and the
value obtained here is $0.61 \pm 0.08$.  This is remarkably close
to values obtained with more traditional approaches. 

The energy density and equation of state of the dark energy may
be obtained as functions of redshift if the value of $\Omega_{0m}$ 
is known.  These were determined assuming $\Omega_{0m}=0.3$. 
The potential energy $V(z)$ is flat, as expected if the dark energy
is a cosmological constant, and $K(z)$ is very close to zero, also as
expected in a LCDM model.  In fact, all 
of the results obtained here are consistent with expectations in 
a standard LCDM model since $P, f, w, K$, and $V$ all remain 
essentially constant
for the redshift range for which the data allow a determination of
each quantity. The two modified gravity models shown in Fig. [\ref{E248}]
and Fig. [\ref{E248diff}]  
do not describe the data as well as the standard LCDM model.  
 
\begin{acknowledgments}
It is a pleasure to thank Paul Frampton for helpful discussions
related to this work. This work was supported in part by the U. S. National
Science Foundation (NSF) under grants AST-0206002, AST-0507465, and by
Penn State University (R. A. D.), and 
by the NSF grant AST-0407448 and the Ajax Foundation (S. G. D.).
We acknowledge the outstanding work and efforts of many observers who
obtained the valuable data used in this study.

\end{acknowledgments}

\bibliography{apssamp}

\end{document}